\documentclass{chi-ext}

\copyrightinfo{
  Copyright is held by the author/owner(s).\\
}

\title{Assessing the Value of 3D Reconstruction in Building Construction}

\numberofauthors{4}
\author{
  \alignauthor{
  	\textbf{Uma Murthy}\\
  	\affaddr{URC Ventures Inc.}\\
  	\affaddr{8201 164th Ave NE}\\
  	\affaddr{Suite 200}\\
  	\affaddr{Redmond, WA 98052 USA}\\
  	\email{uma.murthy@urcventures.com}
  }
  \alignauthor{
  	\textbf{David Boardman}\\
  	\affaddr{URC Ventures Inc.}\\
  	\affaddr{8201 164th Ave NE}\\
  	\affaddr{Suite 200}\\
  	\affaddr{Redmond, WA 98052 USA}\\
  	\email{david.boardman@urcventures.com}
  }
\\
\\
\\
 \alignauthor{
  	\textbf{Chirag Garg}\\
  	\affaddr{Dept. of Building Construction}\\
  	\affaddr{Purdue Univ.}\\
  	\affaddr{West Lafayette, IN 47907 USA}\\
  	\email{cgarg@purdue.edu}
  }
}

\def\plaintitle{Assessing the Value of 3D Reconstruction in Building Construction}
\def\plainauthor{Uma Murthy}
\def\plainkeywords{3D reconstruction, images, user study, building construction}
\def\plaingeneralterms{Human Factors}

\hypersetup{
  pdftitle={\plaintitle},
  pdfauthor={\plainauthor},  
  pdfkeywords={\plainkeywords},
  pdfsubject={\plaingeneralterms},
}

\usepackage{graphicx} % for EPS use the graphics package instead
\usepackage{balance}  % useful for balancing the last columns

\begin{document}

\maketitle

\begin{abstract}
3-dimensional (3D) reconstruction is an emerging field in image processing and computer vision that aims to create 3D visualizations/ models of objects/ scenes from image sets. However, its commercial applications and benefits are yet to be fully explored. In this paper, we describe ongoing work towards assessing the value of 3D reconstruction in the building construction domain. We present preliminary results from a user study, where our objective is to understand the use of visual information in building construction in order to determine problems with the use of visual information and identify potential benefits and scenarios for the use of 3D reconstruction. %Our next step is to design a controlled study to evaluate the effectiveness of 3D reconstruction in comparison with other types of visual information in building construction, such as photos and 3D models from laser scanning. 
\end{abstract}


. 
%See \cite{ACMCCS} for help using the ACM Classification system.


% =============================================================================
\section{Introduction}
% =============================================================================
\begin{figure}
  \centering
  \includegraphics[width=\linewidth]{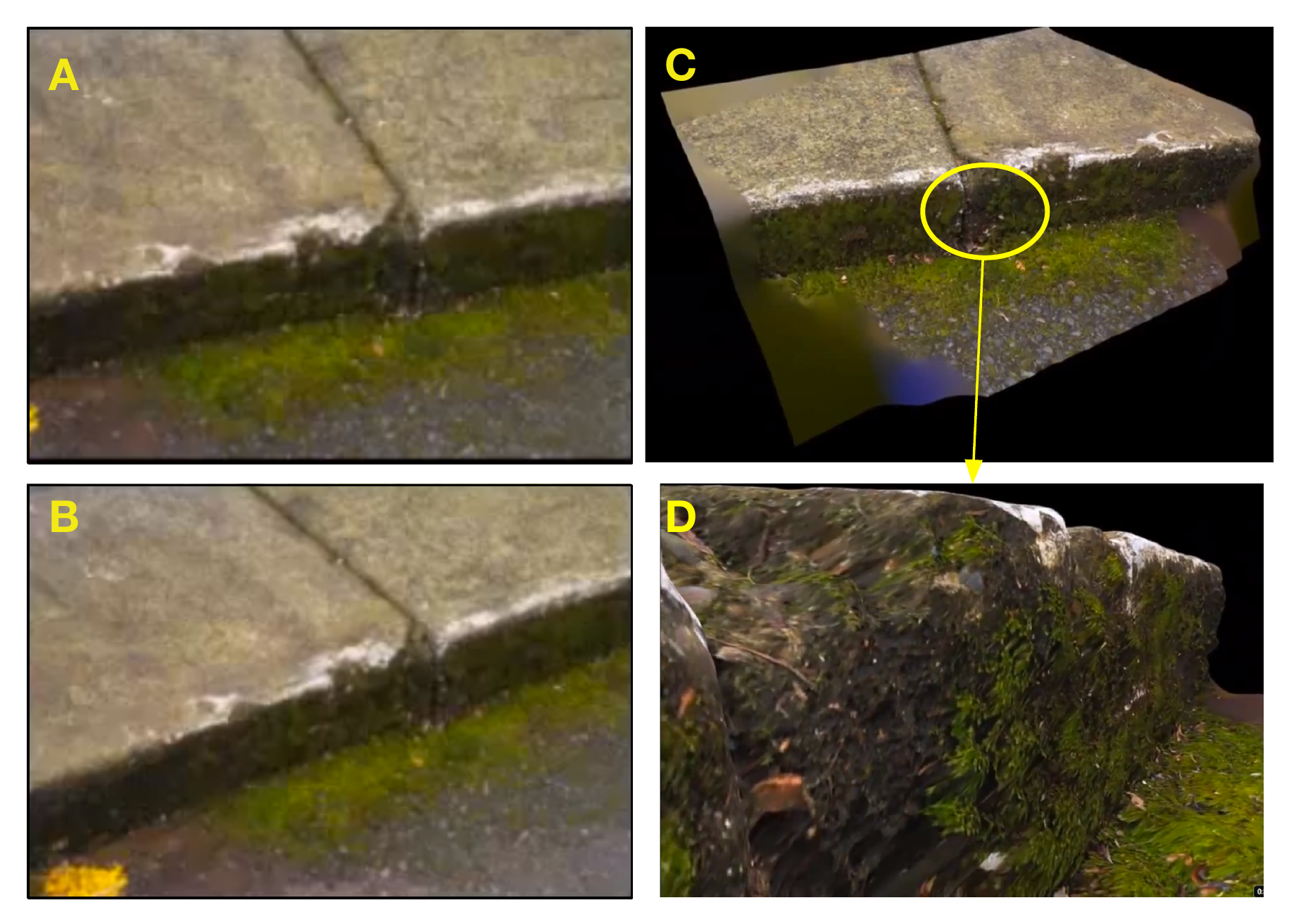}
  \caption{3D reconstruction of a curb: A \& B) Input images of the curb; C) 3D reconstruction model (10 million points) of the curb, zoomed out to reveal the context of the scene; and D) 3D view of the curb, zoomed in to show detail of the crevice between the two stones on the curb.}
  \label{fig:3dr-ex-curb}
\end{figure}
3-dimensional (3D) reconstruction is an emerging field in image processing and computer vision that aims to create 3D visualizations/ models of objects/ scenes from image sets \cite{pollefeys:2000aa,snavely:2006aa}. A set of images is processed to generate a point cloud depicting the real-world scene (as present in the images). This interactive point cloud can be manually processed and manipulated as a 3D model. Figure \ref{fig:3dr-ex-curb} shows the 3D reconstruction (C and D) of a curb from two images (A and B). 3D reconstruction technology has been demonstrated to be effective in military training, tele-operation of vehicles \cite{kelly:2011aa}, tourism \cite{snavely:2006aa}, creating and using story narratives \cite{adabala:2010aa}, and street navigation \cite{kopf:2010aa}. Yet, its commercial applications are yet to be fully explored. In this paper, we describe ongoing work towards assessing the value of 3D reconstruction in the building construction domain. Specifically, we are examining two questions:

1. Can 3D reconstruction from images increase the value of images used in existing construction use cases? \\
2. Can 3D reconstruction from photos provide a comparable alternative to laser scanning (LIDAR) in building construction? \\

To understand the domain and our findings better, we provide a brief overview of the main people involved in a typical construction project. A construction project might involve the following people \footnote{This set of people might vary based on the type of construction project and might include other people, such as a soil engineer or a structural engineer. Also, a group of people might work in the same team -- e.g., the owner and the architect together might be considered a client.}:  \textit{owner} provides the requirements of the construction project; \textit{architect} develops a design from the requirements, including drawings; \textit{general contractor} is in charge of the overall construction of the building and is responsible for converting the design to a constructed building; and \textit{subcontractors}, are responsible for specialized part of the construction, such as masonry or MEP systems engineer. 

There has been prior work researching the use of photos in building construction. PhotoScope \cite{wu:2009aa} provides spatiotemporal visualization of photos to support typical photo seeking tasks in construction, such as for claims and document management. The authors found that presenting the context of time and space facilitates efficient photo searching. Liu and Jones' study on the use of digital photos in the construction industry \cite{liu:2008aa} was aimed at understanding how photos are acquired, stored, edited, viewed, managed, and retrieved. One outcome of their study was a digital image shooting guide to help with better storage, management, and retrieval of photos in building construction activities. 

\marginpar{
Guiding questions for the semi-structured interview:\\

\begin{itemize}\compresslist
\item[1.] Describe your job, your responsibilities, and the activities that you are involved in.
\item[2.] Give examples of contexts/scenarios in which you use visual information.
\item[3.] What problems do you face when you interact with visual information? How do you deal with them?
\item[4.] Give examples of instances when the visual information you had was not sufficient. How did you address this situation?
\item[5.] What contexts/ scenarios do you make site visits?
\item[6.] Have you used LASER technology to develop a 3D model in your job? Can you give examples of those situations.
\end{itemize}
} 
The focus of our studies is to understand the value of 3D reconstruction models (to the user in a construction task) in comparison with other ``visual information'' in building construction. We use the term ``visual information" to represent image-specific information, such as that in a photograph. Visual information in a construction project might include: 1) paper-based and electronic drawings (e.g., plans, detailing, isometric views, and elevations); 2) photos, including those of the construction site, materials, structures, and parts; 3) BIM (Building Information Modeling) models (e.g. REVIT\copyright\footnote{http://usa.autodesk.com/revit-architecture/}) and other 3D models; 4) videos of construction site; and 5) textual description (of a visual), which might be present in construction project documentation, such as a Request-for-information (RFI). 

Most prior work on evaluation of 3D reconstruction technology focused on the evaluation of system performance (efficiency and accuracy), usability of system and method (and not its value), or user performance on a domain-specific task. In their study that compared the (user) performance of a 3D tele-operation approach (a 3D visualization) with video-based tele-operation and direct driving, Huber, Kelly, et al. \cite{kelly:2011aa}. %Participants were asked to drive (or tele-operate) a vehicle on a predetermined route through a challenging obstacle course including several narrow gates, sharp turns, and lane-changes. 
found that their 3D tele-operation approach significantly improved performance, both in terms of driving speed and reduced number of errors when compared to video-based tele-operation. Users reported that they preferred the 3D-based tele-operation mode to the video-based mode as it provided a wider field of view and had the ability to view a scene from arbitrary viewpoints. Overall workload was measured least for the 3D video interface versus the live video and manual drive.

%3D reconstruction models provide a cheaper, simpler, and more efficient alternative to laser-scanned point clouds, an emerging method method used in building construction to model real-world surroundings in 3D. However, it still needs to be determined if 3D reconstruction models provide a comparable alternative to laser-scanned point clouds. 
% =============================================================================
\section{Methods}
% =============================================================================
Our first step was to understand how people in building construction interact with and use visual information. Our objectives were to: 1) identify problems in this interaction/ use and 2) understand potential benefits and scenarios for the use of 3D reconstruction technology. To this end, we have been conducting semi-structured interviews with building construction personnel to learn about their professional activities, especially those that involve the use of visual information. The guiding questions of the semi-structured interviews are listed in the margin. At the time of writing the paper, we conducted and analyzed two interviews (P1 and P2). Each interview lasted about an hour. Also, participants shared materials to support their responses. We analyzed the interviews and materials using open coding while keeping in mind the aforementioned objectives. Section \ref{sec:findings} presents the preliminary findings of the study.

Following this study, our plan is to test selected scenarios with 3D reconstruction models and determine potential benefits. Finally, we would like to validate the benefits of 3D reconstruction by comparing its use with other visual information in the building construction domain as well as identify emergent use/ behavior. 

%\textbf{People in a construction project:} %Before reviewing the findings, it is important to have an overview of the stakeholders/ parties and the communication flow in a building construction project. 
\section{Preliminary findings}
\label{sec:findings}
%We have interviewed two construction professionals so far. 
Participant P1 is an estimator with a construction company that specializes in renovation projects. His job is to review the requirements, design, and the site of a proposed project and estimate the detailed tasks, budget, and timeline for the project. Through this process, he constantly communicates with the project owner (or architect) and with subcontractors and engineers to develop an as accurate as possible budget and schedule estimate. Participant P2 is a project manager in a large construction company that does a variety of projects. His responsibilities include evaluating the design proposed by the architect/ owner, developing a budget and schedule, managing construction personnel on the project, overseeing the construction through completion, managing RFI's and claims. P2's job involves extensive interaction with the owner/ architect, subcontractors, engineers, and other personnel. We present preliminary findings considering our study objectives.

\begin{figure}
  \centering
  \includegraphics[width=\linewidth]{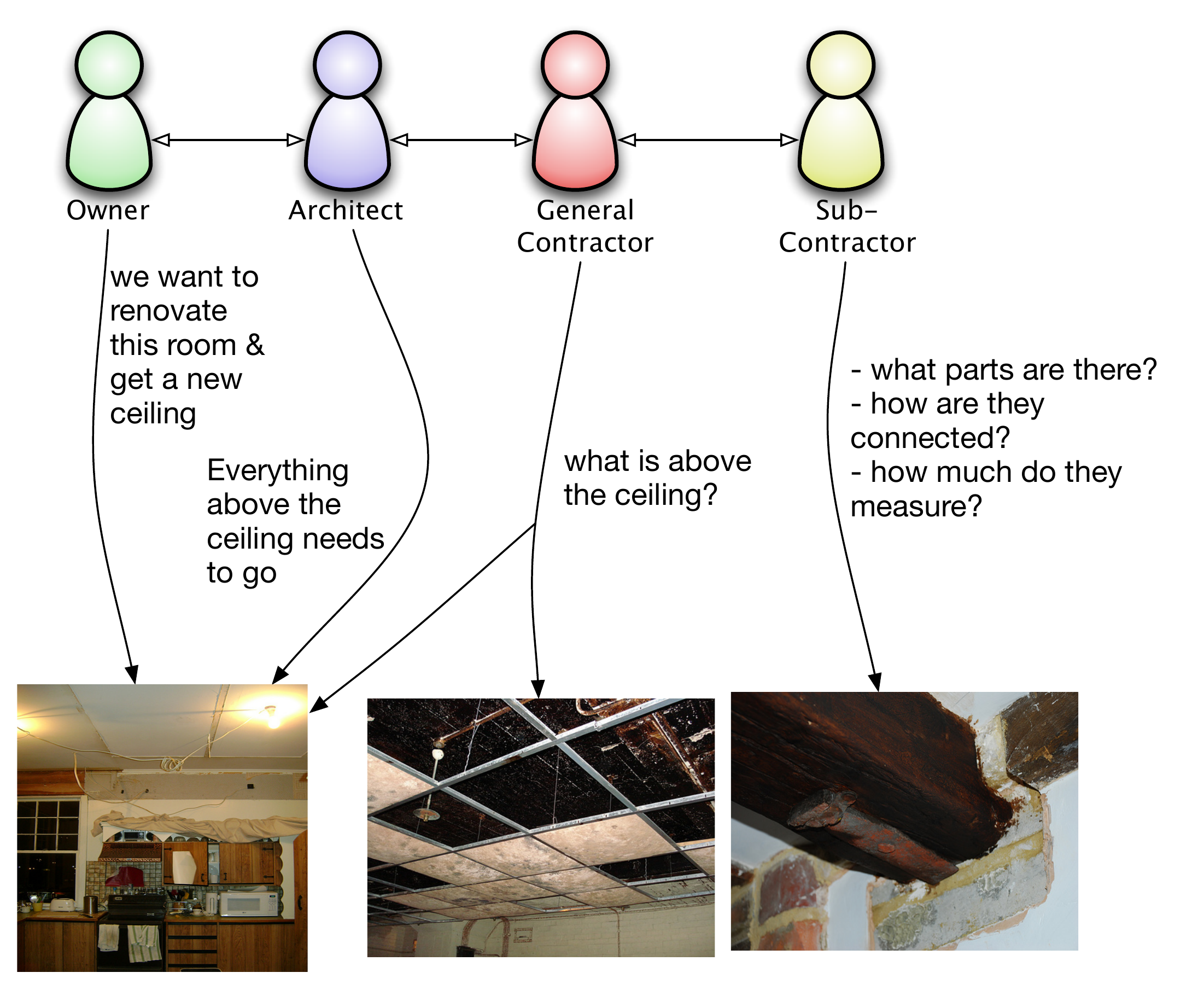}
  \caption{Communication among people in a construction project.}
  \label{fig:comm-flow}
\end{figure}

\subsection{Problems with the use of visual information}
We analyzed participant responses and for problems and found at least three causes:

\textbf{Miscommunication of visual information.} Every person involved in a construction project has their own understanding/perspective of the project design and site. This difference in understanding can lead to miscommunication of specifications, which in turn can lead to increased workload, increased cost, and delays. One example of this communication dynamic is provided in Figure \ref{fig:comm-flow}. %The contractor might not understand In this case, because each person sees 

\textbf{Incomplete or incorrect visual information.} Since each stakeholder in a construction project has their own understanding/ perspective, they might not have all the information as required by the others. In Figure \ref{fig:comm-flow}, note how the need for detail keeps increasing as information flows from the owner through the sub-contractors. The following example from P1 is another instance of incomplete information.

\textit{``We had to renovate a historic building, [which was] constructed in the 1950s. We received the original 30 design drawings and photos as part of the project documentation. If the building had been constructed today, we would have had 30,000 drawings''}

%\marginpar{
%\begin{figure}
%	\centering
%	\includegraphics[width=1.1\linewidth]{../figures/construction-communication-2.pdf}
%	  \caption{Communication among people in a construction project.}
%	  \label{fig:comm-flow}
%\end{figure}
%}

Both participants agreed that in many cases, the design does not reflect the field conditions accurately, leading to incomplete or incorrect visual information. In one example from P2, an incorrect detail of 20 large skylights proved costly for the contractor, resulting in hours of pre-planning and preparation going waste. 

\textit{``The size of the skylight was shorter than the actual concrete opening. ... At the design time, they probably had the skylight hanging outside beyond the concrete opening. But, it ended up being a shorter dimension skylight and getting the water inside the building. We replaced all those skylights with new skylights''}

\textbf{Lack of a sense of orientation in current visual information.} Photographs and drawings are not adequate to understand the the site conditions and orient oneself to the surroundings. For example, a contractor might want to know the locations and dimensions of trees on the site or of neighboring plots and buildings. She might use Google Earth\copyright \footnote{http://www.google.com/earth/index.html} to get a view of the site. However, in most cases that information is not updated.

\marginpar{
\begin{figure}
  \begin{center}
	\includegraphics[width=\linewidth]{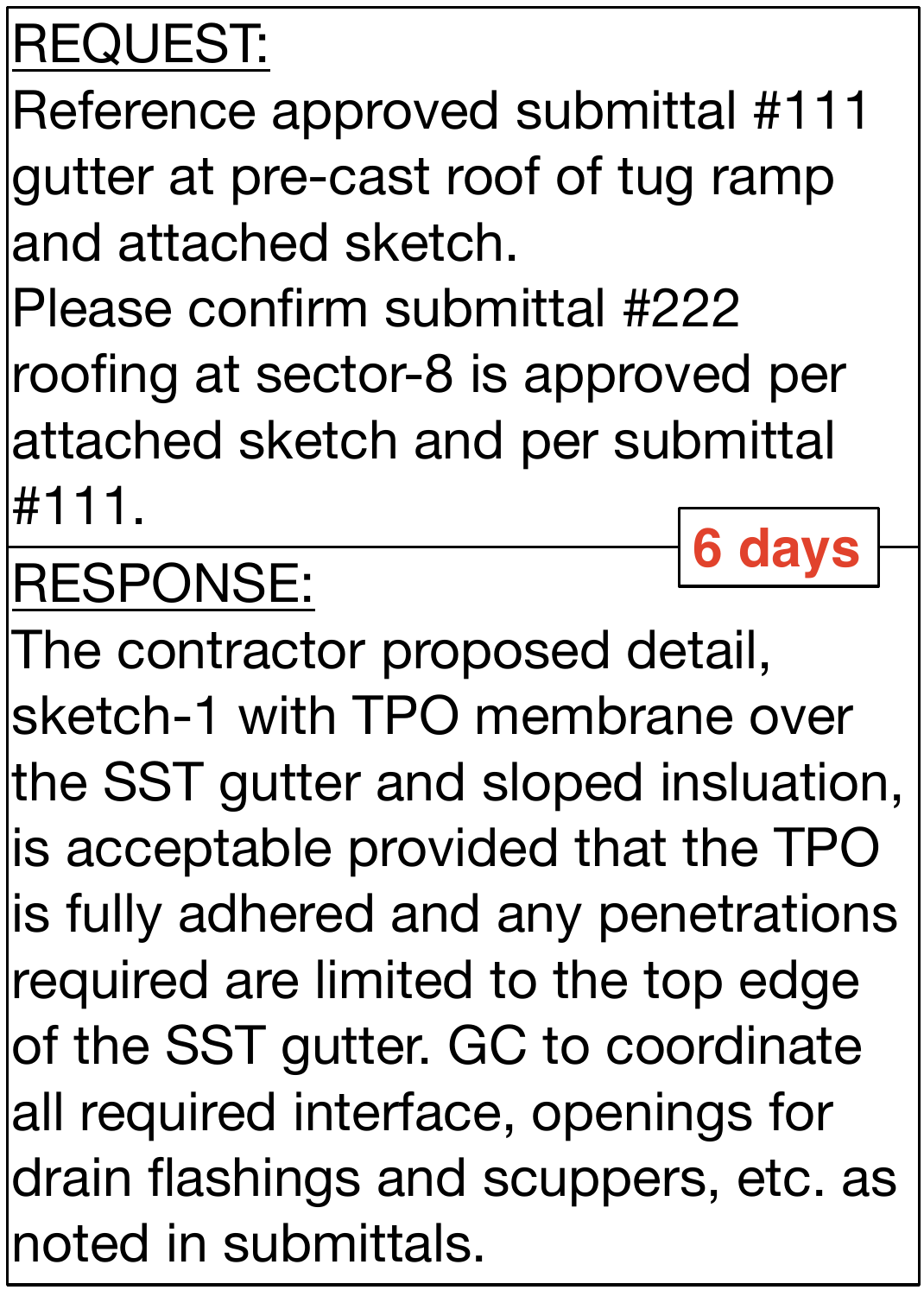}
  \caption{The request, response, and response time in an example RFI document, involving the request for approval of changes made in construction.}
  \label{fig:rfi1}
  \end{center}  
\end{figure}
}
Construction personnel work around these problems in a number of ways. They might engage in frequent formal correspondence to clarify issues, such as exchanging RFIs (see Figure \ref{fig:rfi1}). New visual information and documentation, such as detailed photos, drawings, and 3D models, would need to be developed for parts discovered and to clarify issues. In some cases, parts of the project might need to be re-designed to match field conditions. Related to this is the re-fabrication of parts to fit actual specifications of the site. In most cases, designers, owners, and construction personnel would need to make multiple site visits to orient themselves to the site, to confirm details of parts, to measure (and re-measure) areas for off-site fabrication of parts and to get a complete understanding of the project. All this additional work often would, in turn, result in inefficiency (wasted time), increased costs and workload, and reduced productivity. 

\subsection{Potential benefits and scenarios}
Considering prior work in 3D visualization and modeling, we believe 3D reconstruction has several visualization benefits, most notably being: the ability to experience real-world presence through a sense of \textit{orientation and immersion}; the ability to \textit{view a scene from arbitrary points} to help understand various perspectives of the scene; and \textit{interactive contextual browsing} to understand the context, details, and scale (via geo-referenced points) of a scene. We now outline select scenarios where these capabilities will be beneficial.

Renovation work on an acoustic ceiling (such as that in Figure \ref{fig:comm-flow}): A 3D reconstruction model (among other uses) could be used to communicate and understand the site details, avoid measurement errors, document progress and compare stages, and access context and details of individual parts.\\
Orientation on a construction site: A 3D reconstruction model would facilitate site orientation and understanding (provide a sense of immersion and ability to understand and measure details) to help prepare with construction activities (what is the level is the soil, is there a stone strata to support the foundation), thus avoiding several site visits. \\
Help in pre-fabrication of parts (such as that mentioned in ``skylight'' example): With a 3D reconstruction model, one would have actual measurements of parts of the construction site. By referencing the model, one can avoid mistakes in fabrication (see skylight example above) and prepare parts to match the site specifications accurately. \\
%Serve as supporting visual information: Using a 3D reconstruction model (with geo-referenced points), one would minimize the need for new design, drawings, photos, and other visual information to support project communication and documentation (such as RFIs).

% =============================================================================
\section{Discussion and next steps}
% =============================================================================
\label{sec:discussion}
A 3D reconstruction model can provide rich information for many applications where spatial contexts and interactivity are relevant. In this paper, we presented preliminary findings from the first of a set of user studies to assess the value of 3D reconstruction in the building construction domain. Problems in the use of use of visual information might be caused by miscommunication, incorrect and incomplete information, and lack of orientation. %, work-around strategies to overcome these problems, potential benefits and scenarios By analyzing Through literature review and We saw that We listed some of the problems Our goal is to make a valid case for the need for 3D reconstruction in the building construction domain. 

In the next phase of the project, we will compare the use of 3D reconstruction models with other visual information in construction activities. The scenarios developed from the current study will help in designing our next evaluation and ground the evaluation/experiment tasks in real-world activities. Eventually, our goal is to be able to measure the benefits of 3D reconstruction, irrespective of the domain in which it is deployed. Bowman and Macmahan have a similar discussion on the benefits of immersion \cite{bowman:2007aa}. They describe studies to evaluate the benefits of immersion by isolating its components, such as field-of-view and field-of-regard and examine how each components effects a user's task performance. We would like to explore similar questions in the context of 3D reconstruction, such as what are the objectively measurable components of 3D reconstruction model and how do they impact the performance %(time-to-task, accuracy, and workload) 
of the user? %For example, what are the components of interactive contextual browsing? 
What levels of zooming yield optimal visibility and understanding to support the user's task?

\section{Acknowledgements}
We thank our participants for their valuable inputs. Thanks to Dr. Hazar Dib, for his support in this work. Photo sources for Figure \ref{fig:comm-flow} are from Flickr: \textit{Kitchen renovation} by gmclean, \textit{Dollis Hill bunker, false ceiling tiles} by RachelH\_, and \textit{DSC\_4446} by gordan\_mullan; The RFI information in Figure \ref{fig:rfi1} has been provided by the participants. 

\balance
\bibliographystyle{abbrv}
\bibliography{threedrchiwip}

\end{document}